\documentclass[usenatbib]{mnras}

\usepackage[T1]{fontenc}
\usepackage{ae,aecompl}
\usepackage{graphicx}	
\usepackage{amsmath}	
\usepackage{amssymb}	

\usepackage{url}
\title[Binary BH merger rates inferred from the ULX LF]{Binary black hole merger rates inferred from luminosity function of ultra-luminous X-ray sources}

\author[Y. Inoue et al.]{
Yoshiyuki Inoue,$^{1}$\thanks{E-mail: yinoue@astro.isas.jaxa.jp (YI)}
Yasuyuki T. Tanaka,$^{2}$
and Naoki Isobe$^{1,3}$
\\
$^{1}$Institute of Space and Astronautical Science JAXA, 3-1-1 Yoshinodai, Chuo-ku, Sagamihara, Kanagawa 252-5210, Japan\\
$^{2}$Hiroshima Astrophysical Science Center, Hiroshima University, 1-3-1 Kagamiyama, Higashi-Hiroshima, Hiroshima 739-8526, Japan\\
$^{3}$School of Science, Tokyo Institute of Technology, 2-12-1 Ookayama, Meguro, Tokyo 152-8551, Japan
}

\date{Accepted XXX. Received YYY; in original form ZZZ}

\pubyear{2015}

\begin{document}
\label{firstpage}
\pagerange{\pageref{firstpage}--\pageref{lastpage}}
\maketitle

\begin{abstract}
The Advanced Laser Interferometer Gravitational-Wave Observatory (aLIGO) has detected direct signals of gravitational waves (GWs) from GW150914. The event was a merger of binary black holes whose masses are $36^{+5}_{-4}M_{\odot}$ and $29^{+4}_{-4}M_{\odot}$. Such binary systems are expected to be directly evolved from stellar binary systems or formed by dynamical interactions of black holes in dense stellar environments. Here we derive the binary black hole merger rate based on the nearby ultra-luminous X-ray source (ULX) luminosity function (LF) under the assumption that binary black holes evolve through X-ray emitting phases. We obtain the binary black hole merger rate as $5.8 ({t}_{\rm ULX}/{0.1 \ \rm Myr})^{-1} \lambda^{-0.6} \exp{(-0.30\lambda)} \ {\rm Gpc^{-3}\ yr^{-1}}$, where $t_{\rm ULX}$ is the typical duration of the ULX phase and $\lambda$ is the Eddington ratio in luminosity. This is coincident with the event rate inferred from the detection of GW150914 as well as the predictions based on binary population synthesis models. Although we are currently unable to constrain the Eddington ratio of ULXs in luminosity due to the uncertainties of our models and measured binary black hole merger event rates, further X-ray and GW data will allow us to narrow down the range of the Eddington ratios of ULXs. We also find the cumulative merger rate for the mass range of $5M_\odot\le M_{\rm BH}\le100M_\odot$ inferred from the ULX LF is consistent with that estimated by the aLIGO collaboration considering various astrophysical conditions such as the mass function of black holes.
\end{abstract}


\section{Introduction}
\label{intro}
One hundred years ago, Albert Einstein predicted the existence of gravitational waves \citep[GW;][]{ein16}. Their existence is indirectly established by the discovery of the binary pulsar system PSR~B1913+16 \citep{hul75} and the measurement of  its orbital period derivative \citep{tay82}. 

The twin Advanced Laser Interferometer Gravitational-Wave Observatory (aLIGO) detectors, Michelson-based interferometers with 4~km long arms, \citep{aas15,abb16_ligo} have finally directly detected a gravitational-wave signal from GW150914 very recently \citep{abb16_GW150914}. The event was the merger of a pair of black holes (BHs) whose masses are $36^{+5}_{-4}M_{\odot}$ and $29^{+4}_{-4}M_{\odot}$ and formed $62^{+4}_{-4}M_\odot$ BH with $3.0^{+0.5}_{-0.5}M_\odot c^2$ radiated in gravitational waves. The existence of binary stellar-mass black hole systems is established by this aLIGO observation for the first time. The source locates at a luminosity distance of $410^{+160}_{-180}$~Mpc ($z=0.09^{+0.03}_{-0.04}$). Furthermore, a possible electromagnetic counterpart of the GW150914 event above 50~keV, 0.4~sec after the GW event was detected, is reported by the {\it Fermi} Gamma-ray Burst Monitor (\citet{con16}, but see also \citet{lyu16}). If the association is real, the association indicates that, at least, abundant gas should harbor around the BBHs.

Assuming that the binary BH (BBH) merger rate is constant with in the volume and that GW150914 is representative of the underlying BBH population, the BBH merger rate is inferred to be 14$^{+39}_{-12}$~Gpc$^{-3}$yr$^{-1}$ (comoving frame) at the 90\% credible level \citep{abb16_rate}. Incorporating all triggers that pass the search threshold while accounting for the uncertainty in the astrophysical origin of each trigger, the BBH merger rate is inferred as 6--400~Gpc$^{-3}$yr$^{-1}$ considering various astrophysical conditions such as the spin parameter and the BH mass distributions.

A question is how the progenitor system of GW150914 is formed. Considering the masses of two BHs, the stellar origin of GW150914 is likely to be formed in a low-metallicity environment below $\sim1/2Z_\odot$ \citep{bel10,map13,spe15,abb16_astro}. There are several possible channels to form a BBH such as the isolated stellar binary system scenario \citep[see e.g.][]{tut93,kal07,pos14,kin14,men14,dom15,man16,bel16}, the binary trapped in the active galactic nucleus disc \citep{bar16,sto16}, and the dynamical formation scenario in the dense stellar environment  \citep[see e.g.][]{por00,ole06,ole07,sad08,rod15,ole16}.

Theoretical population synthesis models for isolated binary systems have predicted a wide range of the BBH merger rate as 0--1000~Gpc$^{-3}$yr$^{-1}$ \citep[see e.g.][]{kal07,men14,dom15,man16,bel16}. As massive stellar binary systems would evolve through the X-ray emitting binary phase \citep{rap05,wik15}, BBHs formed from a massive stellar binary system may be evolved through X-ray luminous phase. However, it is currently unclear from both observational and theoretical viewpoints how many fractions binary BHs evolve through X-ray emitting phases. 

Galactic X-ray emitting binaries are known to be as luminous as $\sim10^{38}$~erg s$^{-1}$ and their BH masses are $\lesssim 10M_\odot$ \citep[e.g.][]{fen04} which is lower than the masses of the progenitors of GW150914. Here, in nearby galaxies, there is a more luminous population called as ultra-luminous X-ray sources (ULXs) whose X-ray luminosities is greater than $10^{39}$~erg s$^{-1}$ and locates at off-nucleus positions.  

Two distinct ideas are widely considered to interpret high X-ray luminosities of ULXs, although there is no general agreement on their nature. The first interpretation assumes that they host an intermediate mass black hole with a mass of $M_{\rm BH}\gg 10 M_\odot$ \citep[e.g.][]{mak00}, while the second one invokes a supercritical mass-accretion rate onto a stellar mass black hole\footnote{A ULX powered by an accreting neutron star has been recently discovered by {\it NuSTAR} \citep{bac14}. As the fraction of neutron star powered ULXs is not well understood, we assume that ULXs are powered by BHs in this paper.}. The so-called ultra-luminous state \citep{gla09}, suggested in recent X-ray observations of ULXs \citep[e.g.][]{vie10}, is regarded as a signature of the supercritical accretion rate. However, numbers of theoretical or numerical studies \citep[e.g.][]{vie08} indicated that the X-ray luminosity of the ULXs are not able to exceed a few times of the Eddington luminosity even in the ultra-luminous state due to strong advection and photon trapping within their accretion disc
\citep[e.g.][]{ohs05}. Actually, through detailed X-ray spectral analysis, \citet{iso12} demonstrated that the nature of the ULXs are consistently explained by the scenario that they host a heavy black hole ($M_{\rm BH} \gtrsim 10 M_\odot$) radiating at the trans-Eddington luminosity ($L_X \lesssim L_{\rm Edd}$) but accreting at the supercritical rate. Although BH masses of ULXs are not well constrained due to the situation above, the BH masses of two ULXs have recently been dynamically constrained as $5M_\odot \lesssim M_{\rm BH}\lesssim20-30M_\odot$ for the M101 ULX-1 \citep{liu13} and $3\lesssim M_{\rm BH}\lesssim15M_\odot$ for the ULX P13 in NGC~7793 \citep{mot14}.

ULXs are known to be hosted by low-metallicity galaxies at the metallicity of $Z\lesssim 1/2Z_\odot$ \citep{map10} which is coincident with the local environment for the GW150914 event \citep{abb16_astro}. However, it is known that a galaxy is not chemically homogeneous and can have metallicity dispersion by a factor of 10 in a galaxy \citep[e.g.][]{rol00,nii11}. X-ray measurements of the ULX NGC~1313 X-1 revealed that the local oxygen abundance is $\sim$50\% of the solar value using the low energy X-ray absorption feature \citep{miz07}. Therefore, ULXs are also expected to be formed in the low-metallicity environments like the detected GW event. 

Although the lifetime of ULXs is still not well understood, ULXs are also known to locate at near young stellar clusters \citep[see e.g.][]{gri11,pou13}, which can be used as the age indicator of ULXs. For example, in the colliding star-forming Antennae galaxies, ULXs are associated with the stellar clusters with the age of $<6$~Myr \citep{pou13}.  

The X-ray luminosity function (LF) of ULXs in the local Universe is established with $\gtrsim100$ ULX samples \citep[e.g.][]{gri03,wal11,swa11,min12}. If the massive stellar binaries which become BBHs evolve through an X-ray emitting phase, we can infer the expected BBH merger rate from the ULX LF assuming the duration of the ULX phase and the Eddington ratio of ULX luminosities. 

This paper is organised as follows. In \autoref{sec:ULX}, we introduce the ULX LF and its number density in the local universe. \autoref{sec:rate} presents the expected BBH merger rates inferred from the ULX LF. Discussion and conclusion is given in \autoref{sec:dis} and \autoref{sec:con}, respectively. 

\section{Local ULX Number Density}
\label{sec:ULX}

The X-ray LFs of ULXs have been established in literature \citep[e.g.][]{gri03,wal11,swa11,min12}. However, they are suffered from the completeness of host galaxies in a fixed volume, i.e. the sky completeness. Most of ULX LFs are constructed based on a complete ULX sample in individual galaxies not in a given volume. \citet{swa11} constructed the local ULX LF based on the ULXs which are detected by the {\it Chandra} X-Ray Observatory Advanced CCD Imaging Spectrometer (ACIS). 107 ULXs are identified in a complete sample of 127 nearby galaxies within the volume $V_{\rm ULX}$ of 6100~Mpc$^3$ \citep{swa11}. Since the volume completeness is important to estimate the number density, we adopt the ULX LF of \citet{swa11}. The number density of ULXs per logarithmic luminosity bin in the local universe is given as \citep{swa11}
\begin{equation}
L_X\frac{dn}{dL_X} = \frac{C}{V_{\rm ULX}} \left(\frac{L_X}{10^{39}\ {\rm erg/s}}\right)^{1-\alpha} \exp{\left(-\frac{L_X}{L_c}\right)},
\end{equation}
where $L_X$ is the X-ray band luminosity at 0.3--10~keV, $C=78.0^{+124.5}_{-46.7}$, $\alpha=1.6\pm0.3$, and $L_c=15.2^{+53.0}_{-8.6}\times10^{39}\ {\rm erg\ s^{-1}}$. 

The luminosity function is based on luminosities estimated from X-ray photon counts detected in the {\it Chandra}/ASIC 0.3--6.0~keV band assuming an absorbed power-law spectrum of photon index $\Gamma=1.7$ and a Galactic column density \citep{swa11}. The other luminosity function based on luminosities estimated from spectral model fits is also available, which has $C=26.3^{+9.8}_{-6.8}$, $\alpha=0.8\pm0.2$, and $L_c=16.7^{+20.4}_{-6.8}\times10^{39}\ {\rm erg\ s^{-1}}$. However, a subset of the ULX samples do not have spectroscopic luminosities due to the lack of statistically enough X-ray photon counts for the fit. This can introduce a bias in the LF as discussed in \citet{swa11}. Thus, we use the X-ray photon counts based luminosity function.

BH masses of ULXs are under debate for a long time \citep[e.g.][]{fen11} depending on the assumed Eddington ratio $\lambda$ in luminosities. We can relate the bolometric luminosity to the BH mass as $L_{\rm bol} = \lambda L_{\rm Edd} \simeq 1.26\times10^{38} \lambda {M_{\rm BH}}/{M_{\odot}}\ [{\rm erg/s}]$,
where $M_{\rm BH}$ is the BH mass, $L_{\rm Edd}$ is the Eddington luminosity $4\pi Gm_pcM_{\rm BH}/\sigma_{\rm T}$, $G$ is the gravitational constant, $m_p$ is the proton mass, $c$ is the speed of light, and $\sigma_T$ is the Thomson cross section. As the luminosity is dominated in the X-ray band, we assume $L_{\rm bol}\approx L_X$ \citep[see e.g.][]{fen11} hereinafter. We note that $\lambda$ represents the Eddington ratio in luminosities not in mass accretion rates.

Then, the number density of ULXs per logarithmic BH mass bin in the local universe $n(M_{\rm BH}; \lambda)\equiv M_{\rm BH}{dn}/{dM_{\rm BH}} $ is represented as 
\begin{equation}
n(M_{\rm BH}; \lambda) \simeq 4.4\times10^7 \left(\frac{\lambda M_{\rm BH}}{M_{\odot}}\right)^{-0.6} \exp{\left(-\frac{\lambda M_{\rm BH}}{1.2\times10^2M_{\odot}}\right)}~[{\rm 1/Gpc^3}].
\label{eq:densityall}
\end{equation}

In this paper, we take the value inferred from \autoref{eq:densityall} as the fiducial value. However, it is assumed that all the ULXs have the same Eddington ratio. It is naturally expected that the Eddington ratio of ULXs has a distribution. Observationally, the Eddington ratio distribution function is not well constrained as the exact $\lambda$ of individual sources is still under debate. Here, by assuming a log-normal distribution because of $\lambda>0$, we are able to evaluate how possible distributions affect our results. In this case, the number density of ULXs per logarithmic BH mass bin in the local volume is given by
\begin{equation}
\label{eq:dist}
n(M_{\rm BH}; \lambda)=\int_0^\infty d\lambda' M_{\rm BH}\frac{dn}{dM_{\rm BH}}(M_{\rm BH};\lambda') \frac{dn}{d\lambda'}(\lambda'),
\end{equation}
where the shape of log-normal distribution is $
\frac{dn}{d\lambda'}=\frac{1}{\sqrt{2\pi\sigma_\lambda^2}\lambda'}\exp\left[{-\frac{(\ln\lambda'-\ln\lambda)^2}{2\sigma_\lambda^2}}\right]$, where $\lambda$ and $\sigma_\lambda$ is the location parameter and the scale parameter, respectively. Therefore, the Eddington ratios roughly distribute for the $2\sigma_\lambda$ orders. Since various distribution functions can be expressed by the combination of various log-normal distribution functions, log-normal distributions will allow us to see the possible ranges of the number density of ULXs. 

As GW150914 is the remnant of twin massive stars, we need to consider ULXs in the high-mass X-ray binary (HMXB) systems. It is known that ULXs hosted in spiral galaxies are likely to be in the HMXB systems \citep[see e.g.][]{swa04,swa09,wal11}. The fraction of HMXB systems $f_{\rm HMXB}$ is estimated as $46/82$, since 46 out of 82 ULX host galaxies are spiral galaxies \citep{swa04}. We take this value in this paper. If we assume that all the ULXs in galaxies other than ellipticals are in the HMXB systems, $f_{\rm HMXB}$ becomes 64/82 which increases the inferred merger rate for $\sim$40\%.

The GW150914 event took place inside an almost equal mass binary system as the masses of merged BHs are $36M_\odot$ and $29M_\odot$ \citep{abb16_GW150914}. In our own Galaxy, massive stars are known to be members of binary systems whose mass ratio distribution is flat between $M_1/M_2=0$ and $M_1/M_2=1$ \citep[$M_1$ and $M_2$ is the mass of the companion and primary star in the binary system, respectively;][ ]{kou07,san12,kou14}. The flat mass distribution is confirmed by the HMXB volume density studies \citep{min12}. Following \citet{min12}, we assume that the fraction of the HMXB populations having equal-mass system is $f_{\rm ratio}\sim0.2$. Therefore, the number density of ULXs which are in the equal-mass HMXB system is given as $\rho(M_{\rm BH}; \lambda) = f_{\rm HMXB}f_{\rm ratio}n(M_{\rm BH}; \lambda)$.

As the aLIGO detected the coalescence of a BBH with masses of $36M_\odot$ and $29M_\odot$, the number density of such a system in the fiducial model is $\rho(36M_\odot; \lambda) \simeq 5.8\times10^5 \lambda^{-0.6} \exp{(-0.30\lambda)}~[{\rm Gpc^{-3}}],$ where we adopt the best-fit ULX LF values.

\section{Expected Merger Rate}
\label{sec:rate}
As the number density evaluated in \autoref{eq:densityall} is in the ULX system (a BH + a companion star), we need to take into account the duration of a ULX phase to estimate the expected merger rate. Depending on the stellar metallicities, the progenitor mass of the $29M_\odot$ BH is $\sim35-90M_\odot$ \citep{spe15,abb16_astro}. The companion massive stars heavier than $30M_\odot$ evolve so fast that the duration of a ULX phase becomes shorter than stellar life time of massive stars $\sim$1~Myr. The ULX phase timescale $t_{\rm ULX}$ should be the difference of stellar life times in the binary system which is typically 0.1~Myr for HMXB systems \citep{min12}. 

Observationally, $t_{\rm ULX}$ is hard to be determined. However, ULXs locate to near young stellar clusters \citep[see e.g.][]{gri11,pou13}. Such the association of ULXs with nearby star clusters allows us to evaluate the age of the system by assuming that ULXs are also formed in the same star clusters. For example, in the colliding star-forming Antennae galaxies, ULXs are associated with the stellar clusters with the age of $<6$~Myr \citep{pou13}. In a dwarf irregular galaxy Holmberg~IX, the ULX Holmberg~IX X-1is known to be associated with the stellar cluster with the age of $\lesssim20$~Myr \citep{gri11}. Its optical counterpart star's mass is expected to be $\lesssim15^{+5}_{-5}M_\odot$ \citep{gri11}. Therefore, each ULX must have different timescale. However, its distribution is not well understood. 

In this paper, following \citet{min12}, we take $t_{\rm ULX}$ of 0.1~Myr for all the ULXs as the fiducial value. The expected BBH merger rate inferred from the ULX LF is then approximately estimated as $\dot{\rho}(M_{\rm BH}; \lambda) \approx \frac{\rho(M_{\rm BH}; \lambda)}{t_{\rm ULX}}$.
If we assume shorter $t_{\rm ULX}$, the expected rate will be enhanced by the factor of relative time-scale differences. For the BH mass of $36 M_\odot$, the rate is given as
\begin{equation}
\label{eq:rate}
\dot{\rho}(36M_\odot; \lambda) \simeq 5.8  \left(\frac{t_{\rm ULX}}{0.1\ {\rm Myr}}\right)^{-1} \lambda^{-0.6} \exp{(-0.30\lambda)} \ [{\rm Gpc^{-3}\ yr^{-1}}].
\end{equation}

\begin{figure}
  \includegraphics[width=\columnwidth]{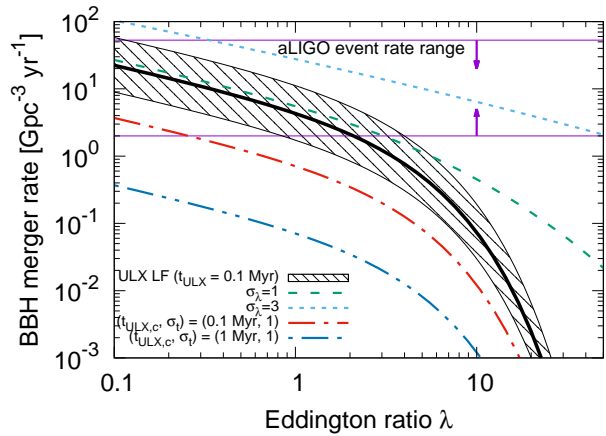} 
\caption{The expected equal-mass ($36M_\odot$ and $29M_\odot$) BBH merger rate as a function of the Eddington ratio $\lambda$ assuming the BBH system is evolved through the X-ray emitting phase. The rate based on the ULX LF \citep{swa11} is shown by the black hatched region where the region is defined by the uncertainty of the normalisation of the ULX LF and the thick solid curve shows the case with the best-fit ULX LF normalisation. We assume $t_{\rm ULX}=0.1$~Myr for all the ULXs. Dashed and dotted curve show the rate considering the log-normal distribution of the Eddington ratio with $\sigma_\lambda=1$ and 3, respectively. For those two curves, $\lambda$ in the plot corresponds to the logarithmic mean value of the ULX Eddington ratio. Dot-dashed and double-dot-dashed curve corresponds to the case of $(t_{\rm ULX,c}, \sigma_t)=$ (0.1~Myr, 1) and (1~Myr, 1), respectively, considering the log-normal distribution of the ULX lifetime but not taking into account the Eddington ratio distribution. The BBH merger rate based on the aLIGO detection 2--53~Gpc$^{-3}$~yr$^{-1}$ for the detected mass is shown by the purple lines with arrows. }\label{fig:rate}
\end{figure}

\autoref{fig:rate} displays the BBH merger rate inferred from the ULX LF of \citet{swa11} assuming that BBHs evolve through X-ray emitting phases. The black shaded region corresponds to the expected merger rate based on the ULX LF assuming $t_{\rm ULX}=0.1$~Myr. The area corresponds to the uncertainty of the normalisation of the ULX LF \citep{swa11} with the thick solid line representing the best-fit one. We do not take into account the uncertainties of the slope and the cut-off luminosity in the ULX LF. Due to the cut-off luminosity $L_c=15.2\times10^{39}$~erg~s$^{-1}$, the expected merger rate exponentially decrease at $\lambda\sim3$. The measured BBH merger rate by the aLIGO collaboration  2--53~Gpc$^{-3}$~yr$^{-1}$ \citep{abb16_rate} is also shown by purple lines and arrows, where it is assumed that all BBH mergers have the same source-frame masses and spins as GW150914. 

Following the fiducial model, the inferred rate is consistent with the measured rate for the range of $0.002\lesssim\lambda\lesssim4$, although $\lambda\lesssim0.2$ does not make the 36 $M_\odot$ BH as a ULX. We note that even the supercritical mass accretion can not exceed a few times of the Eddington luminosity because of strong advection and photon trapping within the disc \citep[e.g.][]{ohs05,vie08}. Therefore, the inferred Eddington ratio range is consistent with the supercritical accretion pictures. 

The dashed and dotted curve show the case with the best-fit ULX LF but considering the log-normal distribution of the Eddington ratio with $\sigma$ of 1 and 3, respectively. The Eddington ratio in the figure corresponds to the logarithmic mean value of $\lambda'$ for a given distribution. $\sigma_\lambda=3$ corresponds to the case in which $\lambda'$ distributes almost uniformly for about 6 orders of magnitude (e.g. at $10^{-3}<\lambda'<10^3$). Therefore, $\sigma_\lambda=3$ would be an extreme case. Even if we consider such an extreme case, the inferred rate from the ULX LF is still consistent with the GW event rate, but at the range of $0.3\lesssim\lambda\lesssim50$.

As discussed above, ULXs are expected to have various lifetimes. Although the distribution of lifetimes is unknown, we can test the case by assuming log-normal distributions as $t_{\rm ULX}>0$.  As in \autoref{eq:dist}, we estimate the rate considering the log-normal distribution for the ULX lifetime. We set $t_{\rm ULX, c}$ as the logarithmic mean of the ULX timescale and $\sigma_t$ as the scale parameter. \autoref{fig:rate} also show the cases of $(t_{\rm ULX,c}, \sigma_t)=$ (0.1~Myr, 1) and (1~Myr, 1) by the dot-dashed and doble-dot-dashed curve, respectively, assuming a log-normal distribution for the ULX lifetime but not taking into account the Eddington ratio distribution. We note again that other possible distributions such as a power-law function can be expressed by the summation of various log-normal distributions. The expected rate is still consistent with the GW event rates but requiring lower Eddington ratios for longer $t_{\rm ULX,c}$. 

Considering various uncertainties, the BBH merger rate inferred from the ULX LF is coincident with the GW event rates reported by the aLIGO collaboration. Uncertainties of our model and the measured BBH merger event rate will not allow us to limit the possible range of $\lambda$ for ULXs under the assumption of ULXs evolving to BBH systems.

\begin{figure}
 \begin{center}
  \includegraphics[width=\columnwidth]{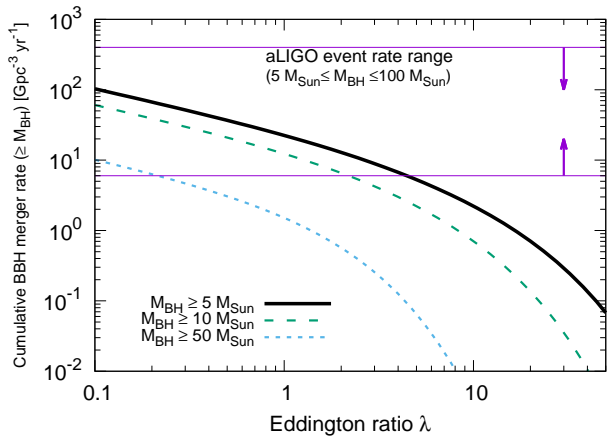} 
 \end{center}
\caption{The expected cumulative BBH merger rates for various mass ranges as a function of the Eddington ratio $\lambda$ assuming the BBH system is evolved through the ULX phase. Solid, dotted, and dot-dashed curve show the rates for $M_{\rm BH}\ge 5M_\odot$, $10M_\odot$, and $50M_\odot$ setting $M_{\rm BH, max}=100 M_\odot$. The best-fit ULX LF of \citet{swa11} is adopted. The distribution of the Eddington ratio is not taken into account. We assume $t_{\rm ULX}=0.1$~Myr. The cumulative BBH merger rate at $5 M_\odot\le M_{\rm BH}\le 100M_\odot$ based on the aLIGO detection 6--400~Gpc$^{-3}$~yr$^{-1}$ considering various astrophysical conditions is shown by the purple lines with arrows. }\label{fig:rate_cum}
\end{figure}

As we adopt the LF of ULXs, the cumulative merger rate above a given BH mass of $M_{\rm BH}$ can be approximately estimated as
\begin{equation}
\dot{\rho}(>M_{\rm BH}; \lambda) \approx \frac{ f_{\rm HMXB}f_{\rm ratio}}{t_{\rm ULX}}\int_{M_{\rm BH}}^{M_{\rm BH, max}} d{M'_{\rm BH}}\frac{dn}{dM'_{\rm BH}}(M'_{\rm BH};\lambda),
\end{equation}
where $M_{\rm BH, max}$ is the maximum mass of the BH and we ignore the distribution of the Eddington ratio.  \autoref{fig:rate_cum} shows the expected cumulative BBH merger rates inferred from the ULX LF for $\ge5$, $\ge10$, and $\ge50 M_\odot$. We use the fiducial parameters and set $M_{\rm BH, max}=100M_\odot$ following \citet{abb16_rate} in which the aLIGO collaboration evaluated the cumulative BBH merger rate for the mass range of $5M_\odot\le M_{\rm BH}\le100M_\odot$ for both BHs with the total mass from 10$M_\odot$ to $100M_\odot$ assuming the distribution of $\propto M_{\rm BH}^{-2.35}$ and a uniform spin parameter distribution. The characteristic frequency $f$ for this mass range will be in the range of $\sim$1--20~kHz as $f$ is given as $c^3/GM_{\rm BH}$ \citep{pea99}. In \autoref{fig:rate_cum}, the cumulative rate estimated using the GW150914 event with the above assumptions is also shown by purple lines and arrows which lie in the range 6--400~Gpc$^{-3}$~yr$^{-1}$\citep{abb16_rate}. The cumulative rate from the ULX LF with $5M_\odot\le M_{\rm BH}\le 100M_\odot$ is consistent with the cumulative rate based on the GW150914 event \citep{abb16_rate}, although the model and measurement uncertainties are still large.

\section{Discussion}
\label{sec:dis}

The coalescence time scale of a BBH due to the dissipation by gravitational waves is \citep{pet64} $
t_{\rm GW} \simeq  3.5\times10^8\left(\frac{a}{20R_\odot}\right)^{4} \left(\frac{M_{\rm BH}}{65 M_\odot}\right)^{-3}~[{\rm yr}]$,
where $a$ is the semi-major axis which is estimated to be $a\gtrsim10-20 R_\odot$ \citep{abb16_astro}. Here, we can roughly compare the inferred $a$ from the GW150914 event with the expected orbital period $P$ of X-ray emitting binary systems. From the Kepler's law,
$P=2\pi \left(\frac{a^3}{G(M_1+M_2)}\right)^{1/2}\simeq1.3 \left(\frac{a}{20R_\odot}\right)^{3/2} \left(\frac{M_1+M_2}{65M_\odot}\right)^{-1/2}\ [{\rm day}]$, where $M_1$ and $M_2$ is the mass of the primary and secondary BH, respectively. The orbital period derived here by assuming $a=20R_\odot$ is consistent with those measured for Galactic HMXBs \citep[0.2--262 days, see e.g.][]{liu11}. Although ULXs are known to be generally highly variable on timescales of days base on a small ULX sample \citep{kaa09,gri10}, the X-ray orbital period of ULXs is not well understood due to its faintness \citep[e.g.][]{gri13}. Several ULXs are known to have orbital periods of $\sim8-60$ days \citep{kaa06,kaa07,liu13,mot14}.

Considering that the GW150914 event is at $z\sim0.09$ and the lifetime of ULXs is in the order of 0.1~Myr, the cosmological evolution effect of the ULX LF will not significantly affect our results as long as we assume that the merger time scale is not cosmologically long, $\lesssim$~Gyr. However, if the parent population of GW150914 is formed at higher redshifts assuming cosmologically long merger time scales, the rate should follow the ULX LF at its formation epoch. 

The cosmic star formation rate density is known to have a peak around $z\sim2$--3 \citep{hop06} which is roughly an order of magnitude higher star formation rate density than in the local universe. Here, at $z\lesssim4$, it is expected that most of stars are formed in the low metallicity environment at $0.02Z_\odot \le Z<Z_\odot$ based on the semi-analytical galaxy formation model \citep[see e.g. Figure. 1 of][]{ino13}. Such low metallicity environment will be suitable for the formation of ULXs and massive BHs. The galaxy formation model can reproduce various observed properties of galaxies such as their LF, luminosity density, and stellar mass density \citep{nag04}, as well as the LFs of high-redshift Lyman-break galaxies and Ly$\alpha$ emitters up to $z\sim6$ \citep{kob07,kob10}. 

As we use the local ULX LF, it is naturally expected that the ULX LF at $z\sim2$ is also about an order of magnitude higher. Therefore, if the progenitor of GW150914 is formed at $z\sim2$, the rate will be enhance by a factor of 10. Due to the uncertainties, that factor still makes the model consistent with the measured rate but requiring higher Eddington ratios.

To form the BBHs as massive as the GW150914 event, the local metallicity environment of $\lesssim1/2Z_\odot$ is required \citep{abb16_astro}. Although ULXs are known to be formed in low-metallicity galaxies at $<1/2Z_\odot$ \citep[e.g.][]{map10} like the GW150914 event, its local metallicity environment is not well understood. \citet{miz07} revealed that the local oxygen abundance of the ULX NGC~1313 X-1 is $\sim$50\% of the solar value using the low energy X-ray absorption feature. Further detailed X-ray spectroscopies of ULXs will allow us to investigate the local metallicity environment of ULXs.

\section{Conclusion}
\label{sec:con}

BBH systems are expected to be directly evolved from stellar binary systems \citep[e.g.][]{tut93,kal07,pos14,kin14,men14,dom15,man16,bel16,sto16} or be formed via dynamical interactions of independent BHs in dense stellar systems \citep[e.g.][]{por00,ole06,ole07,sad08,rod15,ole16,bar16}. If a BBH is evolved from stellar binary systems, it is naturally expected that it passes X-ray emitting phase as many X-ray emitting binaries are observed in nearby universe. In this paper, we have studied the BBH merger rate inferred from the nearby ULX LF \citep{swa11} assuming that BBHs evolve through the X-ray emitting phases. The BBH merger rate is expected to be $5.8 ({t}_{\rm ULX}/{0.1 \ \rm Myr})^{-1} \lambda^{-0.6} \exp{(-0.30\lambda)} \ {\rm Gpc^{-3}\ yr^{-1}}$. Considering various possible channels to form BBHs and stellar binary evolution paths, the inferred rate from the ULXs would represent the subset of the total merger event rates. However, the inferred rate is coincident with the BBH merger rate measured by the aLIGO collaboration at $0.002\lesssim\lambda\lesssim4$ even if we consider the uncertainties of the distribution of the Eddington ratio and the ULX lifetime. 

 Complete ULX LFs with larger ULX samples and accumulation of more GW events by aLIGO \citep{abb16_ligo}, VIRGO \citep{ace15}, KAGRA \citep{aso13}, and LIGO in India will allow us to narrow down the typical Eddington ratio of ULXs which is veiled in mystery. Mass accretion rate can not be simply constrained by this method as the luminosity at supercritical accretion rate is regulated by strong advection and photon trapping within the disc \citep[e.g.][]{ohs05,vie08}.

We also estimate the cumulative BBH merger rate inferred from the ULX LF for $\ge5$, $\ge10$, and $\ge50 M_\odot$. We use the fiducial model parameters and set $M_{\rm BH, max}=100M_\odot$. The inferred rate for $5M_\odot\le M_{\rm BH}\le 100M_\odot$, corresponding to the characteristic frequency of $\sim$1--20~kHz, is consistent with the BBH merger rate estimated by the aLIGO collaboration assuming the same mass range and various astrophysical conditions such as the distribution of spins and masses of BHs. Our results are also consistent with the theoretical predictions of 0--1000~Gpc$^{-3}$yr$^{-1}$ derived by binary evolution population synthesis models \citep[see e.g.][]{kal07,men14,dom15,man16}, although the latter estimation still provide a wide allowed range of event rate. However, it is unclear from both theoretical and observational viewpoints whether BBHs are evolved through X-ray emitting phases. Therefore, further theoretical and observational studies on this coincidence will be necessary.

\bigskip
The authors thank the anonymous referee for his/her careful reading and constructive comments which helped to improve the manuscript. The authors would also like to thank Matteo Guainazzi, Kunihito Ioka, Satoru Katsuda, Bence Kocsis, Kohta Murase, and Tomonori Totani for useful comments and discussions. Y.I. acknowledges support by the JAXA international top young fellowship. YTT is supported by Kakenhi 15K17652.

\end{document}